\documentclass[10pt,conference]{IEEEtran}

\usepackage{balance}
\usepackage{xspace}
\usepackage{shortcuts}
\usepackage{adjustbox}
\usepackage{amsmath}
\usepackage{url}
\usepackage{longtable}
\usepackage{rotating}
\usepackage{afterpage}
\usepackage{graphicx}
\usepackage[many]{tcolorbox}
\usepackage{booktabs}
\usepackage{color}
\usepackage[inline]{enumitem}
\usepackage{comment}
\usepackage{caption}
\usepackage{algorithm,algpseudocode}
\let\oldReturn\Return
\renewcommand{\Return}{\State\oldReturn}
\usepackage{listings}
\lstdefinelanguage{YAML}{
    keywords={name, on, push, jobs, runs-on, steps, uses, with, runs-on, run, id, env, types, matrix, strategy, secrets},
    sensitive=true,
    morecomment=[l]{\#}, 
    morestring=[b]{\"},  
    morestring=[b]{\'},  
    basicstyle=\scriptsize\tt, 
    keywordstyle=\color{blue},  
    stringstyle=\color{red},    
    commentstyle=\color{green}, 
}

\begin{document}

\title{Unlocking Reproducibility: Automating re-Build Process for Open-Source Software}


\author{
	\IEEEauthorblockN{Behnaz Hassanshahi, Trong Nhan Mai, Benjamin Selwyn Smith, Nicholas Allen}
	\IEEEauthorblockA{\textit{Oracle Labs} \\
	Brisbane, Australia \\
	\{behnaz.hassanshahi, trong.nhan.mai, benjamin.selwyn.smith, nicholas.allen\}@oracle.com}
}

\maketitle

\begin{abstract}
Software ecosystems like Maven Central play a crucial role in modern software supply chains by providing repositories for libraries and build plugins. However, the separation between binaries and their corresponding source code in Maven Central presents a significant challenge, particularly when it comes to linking binaries back to their original build environment. This lack of transparency poses security risks, as approximately 84\% of the top 1200 commonly used artifacts are not built using a transparent CI/CD pipeline. Consequently, users must place a significant amount of trust not only in the source code but also in the environment in which these artifacts are built.

Rebuilding software artifacts from source provides a robust solution to improve supply chain security. This approach allows for a deeper review of code, verification of binary-source equivalence, and control over dependencies. However, challenges arise due to variations in build environments, such as JDK versions and build commands, which can lead to build failures. Additionally, ensuring that all dependencies are rebuilt from source across large and complex dependency graphs further complicates the process. In this paper, we introduce an extension to Macaron, an industry-grade open-source supply chain security framework, to automate the rebuilding of Maven artifacts from source. Our approach improves upon existing tools, by offering better performance in source code detection and automating the extraction of build specifications from GitHub Actions workflows. We also present a comprehensive root cause analysis of build failures in Java projects and propose a scalable solution to automate the rebuilding of artifacts, ultimately enhancing security and transparency in the open-source supply chain.

\end{abstract}


\section{Introduction}
\label{intro}

Software ecosystems like Maven are vital to modern software supply chains, offering repositories for
libraries and build plugins. Maven Central, the largest public repository for JVM-based languages,
allows developers to easily reuse existing software and share their own packages.

Unfortunately, Maven Central maintains a separation between binaries and their source code, which results in the lack of a reliable mechanism to link a binary to its corresponding build environment~\cite{trusted-publishers}. Moreover, our investigation reveals that approximately 84\% of the top 1200 most commonly used artifacts on Maven Central are not built using a transparent CI/CD pipeline. This gap poses a significant challenge to enhancing the security of open-source artifacts~\cite{slsa}. As a consequence, users must place considerable trust not only in the source code but also in the environment in which the artifacts are built. However, trust is a highly sensitive and critical issue, especially for organizations that cannot afford to blindly rely on third-party dependencies. For instance, sectors such as government, energy, and other critical infrastructure demand full transparency and control over every single dependency in their systems.

One solution to this problem is rebuilding artifacts from source. This approach enhances supply chain
security for several reasons. First, it allows for a thorough review of the code,
enabling the detection of vulnerabilities, malicious code, or backdoors before compilation. This level of
transparency ensures that only the intended functionality is included.
Second, building from source enables verification that the resulting binaries match the original
source code, significantly reducing the risk of tampering associated with relying on pre-compiled
binaries from untrusted sources. Additionally, this method empowers developers to manage and inspect
all dependencies, ensuring the use of trusted libraries and the exclusion of unwanted or vulnerable
components. Developers can also modify the source code to address specific security needs or disable
potentially risky features, adding an extra layer of protection. Moreover, building from source
allows organizations to maintain compliance with security policies through the auditing of both
the source code and the build process. This is particularly critical in regulated industries that
require adherence to specific standards.

Rebuilding software artifacts from source offers substantial security benefits, yet it also introduces several challenges. Variations in build environments, such as different JDK versions, build commands and tools, as well as source code commits, can lead to build failures. For instance, a recent study~\cite{aroma} showed that 44 out of 100 randomly sampled Maven packages, which included the \codett{project.build.outputTimestamp} property in their POM file (a signal of reproducibility efforts by maintainers), failed to build. This underscores the difficulties of achieving reliable rebuilds in real-world scenarios. While much of the industry focus has been on achieving bitwise reproducibility~\cite{reproducible-builds}, our findings highlight that even ensuring successful rebuilds, let alone bitwise equivalence, at scale is an immensely challenging task. Furthermore, to effectively enhance supply chain security, it is imperative that all dependencies of a project are also built from source. However, dependency graphs are often large and complex~\cite{sbom-paper}, complicating the manual management of the rebuilding process. Therefore, there is a pressing need for a solution that can automate and scale the process of rebuilding artifacts across potentially vast and intricate dependency chains, ensuring both efficiency and effectiveness.

In this paper, we take a significant step towards providing a tooling infrastructure that automates the process of rebuilding artifacts. This automation not only enhances our understanding of the challenges involved but also helps improve the success rate of fully automatic rebuilds. Additionally, we highlight the pressing issues that the community should address to further streamline the process. Specifically, we address the challenge of building open-source Maven artifacts from source and introduce \tool, an extension of Macaron~\cite{macaron}~\footnote{\url{https://github.com/oracle/macaron}}, an industry-grade open-source supply chain security framework. \tool aims to enhance automatic builds while minimizing the need for manual intervention.

As noted in~\cite{aroma}, a key obstacle to automating the rebuilding of Maven packages is obtaining the correct source code. Existing tools~\cite{rc} often assume that the exact version of the source code corresponding to a specific release is readily available, which creates a significant barrier to automation. To address this challenge, we present a new approach compared to AROMA~\cite{aroma}, offering significantly improved performance in automating the detection of source code.

We introduce a novel approach for extracting build commands by analyzing build scripts from upstream projects on GitHub. Building on the build platform check introduced in Macaron~\cite{macaron}, which identifies reachable artifact build and deployment commands, we enhance its capability to extract crucial build information. The original analysis in Macaron constructed a call graph of shell commands invoked by GitHub Actions workflows. Our extension takes this further by incorporating dataflow analysis to resolve GitHub Actions variables and introducing a new abstraction to model third-party GitHub Actions responsible for setting up the build environment. Additionally, we introduce new heuristics to improve the accuracy of detecting build and deployment steps within workflows, coupled with a confidence scoring mechanism to rank these commands. This flexible approach enables us to attempt alternative commands when some fail. Notably, our work also supports Gradle projects, which present distinct challenges compared to Maven-based projects and have been overlooked in prior work~\cite{aroma}.

To ensure that the rebuild-from-source approach is reliable in practice, we need a mechanism to validate the produced artifacts and confirm their equivalence to the original artifacts published on Maven Central. However, without a successful build, there are no artifacts to validate. Additionally, determining equivalence between binaries produced via alternative builds is a separate, orthogonal task, which can be approached using techniques proposed in~\cite{dietrich2024levels}. Previous work~\cite{aroma} focused on bitwise equivalence, and similar studies on binary equivalence validation~\cite{bineq-dataset} have addressed similar challenges. In this work, we focus on improving the automation and success rate of rebuilding Maven artifacts from source, an inherently challenging task in its own right.

In summary, this paper presents the following contributions:

\begin{itemize}
    \item A comprehensive analysis of the transparency of software supply chains through automated artifact release practices in the top 1,200 Maven artifacts.
    \item A novel mechanism for identifying the correct source code for published Maven artifacts, overcoming challenges in detecting code for specific artifact versions.
    \item An automated system for generating build specifications, facilitating the rebuilding of Maven artifacts, supporting both Maven and Gradle build tools.
    \item A detailed root cause analysis of build failures encountered in Java~\footnote{Oracle Java is registered trademarks of Oracle and/or its affiliates. Other names may be trademarks of their respective owners.} projects, identifying common issues and proposing solutions for more reliable rebuilds.
    \item Extending Macaron, an industry-grade, open-source tool with new features to automate the process of rebuilding open-source artifacts, enhancing build automation and security in open-source supply chains.
\end{itemize}

\section{Preliminary Study and Motivation}
\label{sec:motivation}

To understand the current state of transparency of artifact release practices, we examined the release automation among the most frequently used Maven artifacts.\footnote{This experiment was conducted in October 2024.} To carry out this analysis, we constructed a dataset comprising the 1,200 most commonly used unversioned Maven artifacts (identified by their group and artifact IDs, or GAs) from libraries.io~\footnote{https://libraries.io/api, retrieved 03/09/2024.}, excluding two GAs that did not have Maven Central as their package URL. We then retrieved the latest version of each artifact for further analysis.

The primary goal of this study is to address the following key research questions:

\begin{itemize}
    \item What percentage of Maven packages are released before the corresponding code is committed to GitHub?
    \item How often are Continuous Integration (CI) runs deleted, potentially obscuring the release process?
    \item What proportion of packages include provenance metadata, ensuring traceability of the build process?
    \item How many packages are released with observable build deployment in CI pipelines?
\end{itemize}

To answer these questions, we began by identifying the commit associated with each Maven artifact on GitHub. We then compared the artifact's publish timestamp on Maven Central with the corresponding commit timestamp. If the artifact was published before the code was committed, it indicated a lack of automated CI/CD pipeline for the release, making the build process opaque. In such cases, the absence of an observable build process raises concerns about the reproducibility of the artifact.

For the remaining artifacts, where the publish timestamp is later than the commit timestamp, we investigated whether a corresponding CI pipeline existed to trigger the deployment. This would confirm that the artifact was released as part of a properly configured CI/CD process. However, a significant challenge arises from GitHub’s retention policy, which deletes CI run data for public repositories after 90 days~\cite{github-retention-policy}. This policy not only complicates the auditing process but also invalidates provenance data, hindering the ability to track the full history of the artifact and assess its security.\footnote{\url{https://github.com/orgs/community/discussions/138249}} While there have been discussions about updating the retention limit and the upcoming ``Action Immutability'' feature~\footnote{\url{https://github.com/github/roadmap/issues/592}}, it remains unclear how these changes will affect public repositories. Furthermore, even if implemented, these updates will only apply to artifacts produced after the features are enabled, not to those created before.

Table~\ref{tab:release_transparency} summarizes the findings of our study on artifact release transparency. Out of the 1198 artifacts analyzed, we were able to locate the source code for 80.38\% of the artifacts. We were able to find a transparent release pipeline via GitHub Actions for 191 (15.95\%) artifacts. In contrast, 87 artifacts (7.26\%) had code committed after the artifact was released, indicating a lack of CI/CD automation. A significant number of artifacts, 471 (39.31\%), had GitHub Actions run details removed, limiting transparency. Additionally, only 2 artifacts (0.17\%) contained SLSA provenance metadata, which serves as evidence of a secure and verifiable build process. These findings highlight the considerable challenges in achieving full transparency in software supply chains.

Given these challenges, it is clear that a scalable and automated solution for rebuilding open-source projects is necessary. Such a solution would not only enhance transparency but also ensure the reproducibility and reliability of software releases, which are critical to maintaining the security and integrity of the software supply chain.

\begin{table}[ht]
\centering
\resizebox{\columnwidth}{!}{
\begin{tabular}{|l|c|c|}
\hline
\textbf{Finding} & \textbf{Count} & \textbf{Percentage} \\
\hline
Total Artifacts & 1198 & 100\% \\
\hline
Source Commit Found for Artifacts & 963 & 80.38\% \\
\hline
Code Committed After Release & 87 & 7.26\% \\
\hline
GitHub Actions Run Details Removed & 471 & 39.31\% \\
\hline
Release Pipeline via GitHub Actions & 191 & 15.95\% \\
\hline
SLSA Provenance Metadata & 2 & 0.17\% \\
\hline
\end{tabular}
}
\caption{Summary of Artifact Release Transparency Findings}
\label{tab:release_transparency}
\end{table}

\section{Automated Artifact Rebuilding from Source}
\label{design}

To effectively scale the build-from-source process, we require a comprehensive end-to-end solution that can:

\begin{itemize}
    \item Locate the source commit corresponding to the published artifact.

    \item Analyze GitHub Actions build pipeline specifications to identify the build commands and required Java version.

    \item Generate a build specification compatible with build infrastructures, such as Reproducible Central.

    \item Validate the integrity and correctness of the build.
\end{itemize}

In this section, we elaborate on the design of each of these components in detail.

\begin{figure*}
  \begin{center}
    \includegraphics[width=0.7\textwidth]{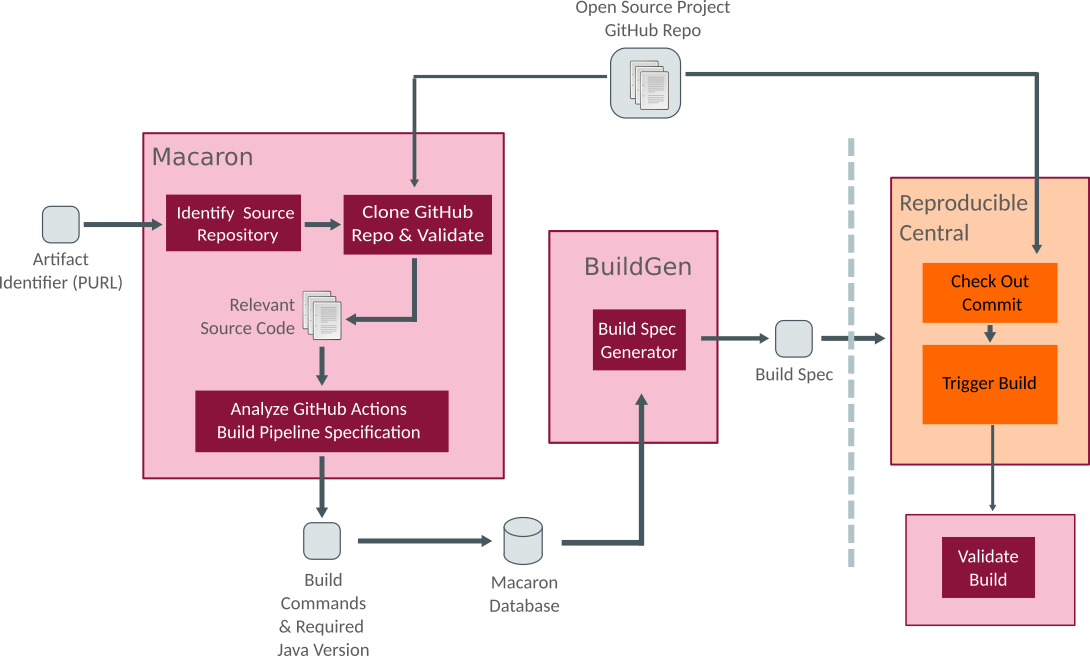}
  \end{center}
  \caption{Build from Source Automation Pipeline}
  \label{fig:arch}
\end{figure*}

\subsection{Finding Source}

To identify the source code repository and the corresponding revision for a given artifact, we developed a component called \commitfinder. First, \commitfinder attempts to find a cryptographically verifiable SLSA provenance~\cite{slsa} for the artifact, which includes information such as the repository and commit hash. The generation of such provenances has recently been supported by major platforms, including GitHub~\cite{github-attestations}. If a provenance is found, \commitfinder directly uses the repository URL and commit hash specified within it. If no provenance is available, \tool retrieves the project link from various fields in the POM file or queries the Open Source Insights service~\cite{depsdev} to gather a set of candidate repository links. These URLs are then normalized and filtered, and \commitfinder attempts to identify a valid repository from the list. Once the repository URL is identified, the matching process uses regular expressions (regex) and a sorting mechanism to find a Git tag that matches the artifact version.

Through extensive preliminary investigations, we identified numerous artifacts with unconventional tag structures used by developers, which influenced the design of the tag-matching methodology in \commitfinder. A critical aspect of this methodology is determining which parts of the repository tag correspond to version strings, as opposed to other elements such as prefixes or suffixes. Incorrect identification of these elements can result in mismatched versions. To address this, we initially established strict rules for handling prefixes and suffixes; however, these rules occasionally failed to match certain artifact versions. To improve this, we relaxed the constraints, enabling more matches but requiring an additional sorting mechanism.

The sorting mechanism in \commitfinder determines the most accurate match when multiple candidates are found. To enhance the accuracy of this step, several improvements were made. These include prioritizing shorter prefixes, selecting prefixes that are substrings of the artifact’s name (as opposed to only considering exact matches), and favoring tags that include significant keywords such as "release." These refinements help improve the match's relevance and reliability.

To mitigate the increased complexity introduced by these improvements, we incorporated a simplified evaluation method as a pre-processing step. This acts as an early termination feature, enabling the system to quickly select tags that directly or nearly match the artifact version without performing the more computationally expensive full evaluation.

\subsection{Detecting Build Information}

In this section, we present our approach to extract build-related information from GitHub Actions workflows. This includes identifying build commands, Java versions, and build tools essential for compiling open-source artifacts. The build information detection uses dataflow analysis, heuristics, and abstractions as illustrated in Algorithm~\ref{alg:build-cmd}. The diverse nature of open-source projects, where build commands can vary across repositories, is accounted for through this approach. By analyzing GitHub Actions workflows in conjunction with repository metadata, the algorithm can extract the necessary information to rebuild an artifact.

The first step in the detection process involves constructing a call graph for the repository, which represents the relationships between functions and their invocations. This graph captures not only the inlined shell scripts but also the transitive analysis of external shell scripts invoked within the GitHub Actions workflows. A breadth-first search (BFS) is then applied to traverse the graph and identify potential build command nodes. Once build commands are detected, the algorithm backtracks through the call graph to identify relevant build configurations, such as environment variables and triggering events. The algorithm also captures publishing configurations (e.g., signing tokens), which influence the build process. Finally, confidence scores are assigned to each detected build command based on predefined heuristics, which estimate the accuracy of each match.

\begin{algorithm}[t]
\caption{Build Information Detection}
\label{alg:build-cmd}  
\begin{algorithmic}[1]
\State \textbf{Input:} $Repo$ \Comment{GitHub repository containing the Actions workflows}
\State \textbf{Output:} $Build\_CMDs$, $Confidence$ \Comment{List of detected build commands and their respective confidence scores}
\State
\Procedure{BuildCommandDetection}{}
    \State \textbf{Step 1:} Construct the call graph $CG$ of the repository using the function \textsc{ConstructCallGraph}($Repo$)
    \State \textbf{Step 2:} Traverse the call graph using breadth-first search (BFS) to identify potential build command nodes
    \State \textbf{Step 3:} For each detected build command node, traverse the caller nodes to identify associated GitHub actions
    \State \textbf{Step 4:} Fetch relevant models or configurations associated with the identified GitHub actions
    \State \textbf{Step 5:} Check if any GitHub variables are used in the identified actions
    \If{GitHub variables are present}
        \State \textbf{Step 6:} Perform backward analysis to resolve the values of the GitHub variables at their declaration point
    \EndIf
    \State \textbf{Step 7:} Collect information regarding the triggering event (e.g., push, pull request, etc.) of the workflow
    \State \textbf{Step 8:} Gather publishing configurations, such as signing tokens, which could impact the build process
    \State \textbf{Step 9:} Assign a confidence score to each identified build command based on heuristics, such as the presence of known keywords or direct matches
    \State \textbf{Step 10:} Return the list of detected build commands, along with their associated confidence scores
\EndProcedure
\end{algorithmic}
\end{algorithm}

\prg{Dataflow Analysis:}  
The analysis begins with constructing a call graph of shell commands invoked by GitHub Actions workflows. This graph helps identify whether build commands are reachable from the workflow configuration. However, it cannot resolve GitHub Actions variables, which are often used to influence the build process. To address this, we use additional dataflow analysis to resolve these variables and understand how they influence the build process.

\prg{Abstraction for Third-Party GitHub Actions:}  
Our system provides abstractions for third-party GitHub Actions, such as \codett{setup-java}, which are commonly used to configure the build environment. These abstractions allow us to identify information, such as the JDK version and distribution required for the build. Since most projects utilize official GitHub Actions hosted by GitHub, the effort required to set up these models is minimal.

\prg{Heuristics for Precision:}  
New heuristics have been introduced to enhance the precision of detecting build and deployment steps within the workflows. These heuristics enable the system to differentiate between various command types, focusing on those directly associated with building and deploying the artifact.

\prg{Assigning Confidence Scores:}  
To further refine the detection process, we propose a confidence scoring mechanism. This allows the system to rank identified build and deployment commands according to their likelihood of being actually used for the target artifact. In cases where certain commands fail, the system can attempt alternative commands.

\begin{lstlisting}[
    language=YAML,
    float,
    mathescape,
    caption={GitHub Actions Workflow Example},
    label={lst:cmd-detection}
  ]
name: Release
on:
  release:
    types: [published]
jobs:
  build:
    runs-on: ubuntu-latest
    strategy:
      matrix:
        java: ['17', '21']
    steps:
      - name: Setup GraalVM CE
        uses: graalvm/setup-graalvm@v1.3.1
        with:
          distribution: 'graalvm'
          java-version: ${{ matrix.java }}
          github-token: ${{ secrets.GITHUB_TOKEN }}
        run: |
          ./gradlew check --no-daemon --continue
  release:
    runs-on: ubuntu-latest
    steps:
      - name: Set up JDK
        uses: actions/setup-java@v4
        with:
          distribution: 'temurin'
          java-version: '17'
      - name: Publish to Sonatype OSSRH
        id: publish
        env:
          GPG_KEY_ID: ${{ secrets.GPG_KEY_ID }}
          GPG_PASSWORD: ${{ secrets.GPG_PASSWORD }}
        run: |
          ./gradlew publishAllPublicationsToBuildRepository publishToSonatype closeAndReleaseSonatypeStagingRepository
          ./gradlew docs
\end{lstlisting}

Consider the GitHub Actions workflow example shown in Listing~\ref{lst:cmd-detection}. In this workflow, there are two jobs: the first job builds the artifact using GraalVM, while the second job is responsible for both building and publishing the artifact to Maven Central. The call graph analysis identifies three potential build-related commands, all invoking \codett{./gradlew}. The analysis then detects the \codett{release} triggering event, identifies the invocation of the \codett{setup-java} GitHub Action, captures the GPG signing key used within the \codett{release} job for artifact publishing to Maven Central, and identifies the Gradle tasks used in each command. Based on the gathered information, the algorithm assigns the highest confidence score to the command on line 32 and reports it as the build command.

\subsection{Build Specification Generation}
\label{subsec:buildspec}

\begin{figure*}
  \begin{center}
    \includegraphics[width=0.6\textwidth]{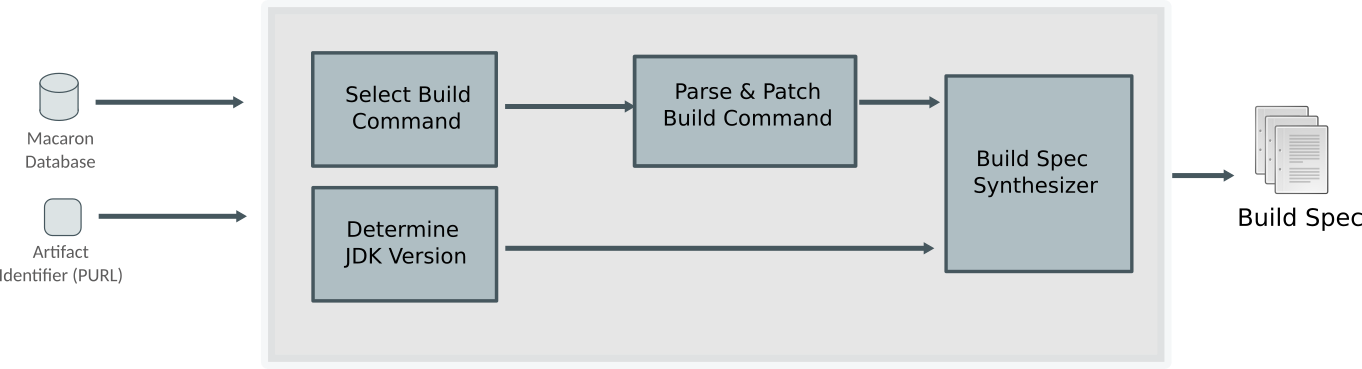}
  \end{center}
  \caption{Workflow for Generating Build Specifications}
  \label{fig:specgen}
\end{figure*}

A requirement for rebuilding artifacts at scale is the ability to formalize and standardize the build process across software components. To address this challenge, we introduce \tool, an extension to Macaron, which automates the generation of build specifications for previously analyzed software artifacts.

A build specification encapsulates all essential metadata required to reproduce the build of a given software artifact, based on the analysis results generated by \tool in earlier stages. This includes key information such as the software’s package coordinates (e.g., group ID, artifact ID, version), the source repository and tag, the build tool used (e.g., Maven or Gradle), the exact build command, the target build output, and the JDK version used for compilation.

The generated specification adheres to the \textit{Reproducible Central Buildspec} format, a structured template specifically designed to support automated rebuild pipelines and facilitate reproducibility validation. By automating the generation of this specification directly from Macaron's internal analysis database, we significantly reduce the manual effort typically required to define build steps.

Figure~\ref{fig:specgen} illustrates the process, which starts with a software component’s PURL (Package URL) and a database containing the results of previous build information collection. \tool then selects the build command with the highest confidence score. If no suitable build command is available, it defaults to a pre-configured build command.

\tool utilizes two command parsers that are aware of Maven and Gradle semantics to parse the selected build command. Once parsed, the build command is processed and patched to account for system properties, task modifications, and other flags required for specific behaviors. This includes actions such as batch processing, skipping tasks, or skipping Javadocs. For example, disabling the \codett{-threads} option was crucial for successfully rebuilding certain artifacts.\bh{Add number of such cases and an example as foornote?}

Simultaneously, \tool identifies the appropriate JDK version. It first attempts to retrieve the version from the GitHub Actions metadata. If unavailable, it falls back to extracting the JDK version directly from the JAR file. The version is then normalized to the major version supported by RC.

Once the build command, JDK version, source repository, and associated commit are determined (as outlined in the previous stage), \tool generates a \textit{Reproducible Central Buildspec} using a predefined template. This structured output can then be directly consumed by rebuild infrastructures, such as Reproducible Central, enabling its integration into larger verification workflows.  Reproducible Central works by taking the generated build specification and using it to automatically create a tailored Docker image with the appropriate JDK and build tools installed. This dynamic approach eliminates the need for pre-configured Docker images, streamlining the process and ensuring that each build environment is correctly set up on-demand, based on the specific requirements of the build specification.

\bh{Add an example build spec}

\subsection{Validating the Build}
\label{subsec:validate_build}

When validating the correctness of a build, our primary objective is to determine whether the build completes successfully and the expected artifacts are generated, rather than ensuring bitwise equivalence between artifacts. While bitwise equivalence may be important in certain contexts, it is not the primary focus in this work. Instead, we concentrate on verifying that the build process executes without errors and that the correct artifact is produced. A successful build is one in which no errors, such as compilation issues or misconfigured dependencies, occur, and the expected artifact is generated as intended.

Rebuilding at scale introduces several challenges, particularly due to variations in build environments, tool configurations, and system resources. These factors can influence build outcomes, with issues like network errors, outdated dependencies, or discrepancies between local environments potentially causing inconsistent results. Consequently, our focus is on improving the build success rate. Determining binary equivalence across alternative builds falls outside the scope of this paper, but can be addressed using the methodologies outlined in~\cite{dietrich2024levels, xiong2022towards, sharma2025canonicalization}.

\section{Evaluation}

We have integrated \commitfinder and \tool as new features within the open-source project Macaron\footnote{https://github.com/oracle/macaron}. The \commitfinder feature is already available in version 0.16.0 of Macaron, while \tool is currently under review. The \spec generated by our tool is fully compatible with Reproducible Central and can be executed on this infrastructure. To evaluate the effectiveness and practicality of our approach, we structure our evaluation around the following research questions.

\begin{enumerate}[label=\textbf{RQ\arabic*:}]
    
    \item \textbf{Source Code Validation:} How reliably can our system identify and validate the source code corresponding to published binary artifacts in software repositories?

    \item \textbf{Build Success on Reproducible Builds:} To what extent can our system successfully reproduce builds for known reproducible projects (RC projects) using the generated build specifications?

    \item \textbf{Build Success in the Wild:} How effective is our system at building newly discovered packages from Maven Central that have no prior build reproducibility metadata?

\end{enumerate}




\subsection{RQ1: Source Code Validation}

In this section, we evaluate \commitfinder's effectiveness in matching source code to binary artifacts by comparing its results against AROMA~\cite{aroma}. This evaluation is conducted using AROMA's dataset, which includes 473,351 artifacts~\footnote{https://zenodo.org/records/10003645}.

We compared the results of \commitfinder with AROMA on a per-artifact basis, ensuring that the detected matches aligned between both tools. Of the 473,351 artifacts analyzed, 198,200 had no possible match due to missing repository URLs or tags. For the remaining artifacts, our results demonstrate that \commitfinder significantly outperforms AROMA in detecting source code commits. Specifically, \commitfinder successfully matched 273,811 artifacts, whereas AROMA achieved 234,674 matches. This represents a clear improvement in \commitfinder's ability to identify relevant source-binary matches. Below are some examples where \commitfinder provided better results than AROMA.
\begin{itemize}
    \item Version \codett{0.25-rc1}\footnote{ai.catboost/catboost-spark-macros\_2.11@0.25-rc1}: 
    \commitfinder found the tag \codett{v0.25}, while AROMA did not find any matching tags because of the suffix in the version.

    \item Version \codett{1.0}\footnote{zone.gryphon.jordan/jordan@1.0}: 
    \commitfinder found the tag \codett{jordan-1.0}, while AROMA did not find any matching tags because of the prefix in the tag.

    \item Version \codett{1.0.1}\footnote{biz.netcentric.filevault.validator.maps/aem-classification-maps@1.0.1}: 
      \commitfinder correctly identified the tag \codett{aem-classification-maps-1.0.1}, whereas AROMA incorrectly reported the tag \codett{1.0.1}. This repository has three distinct \codett{1.0.1} tags with different commit hashes. \commitfinder selects the tag that is most likely the best match based on the GAV (Group, Artifact, Version).
\end{itemize}

To further validate our results, we cross-referenced the matched commits with the source repository contents at the identified commit. A sample set of 71,080 artifacts was selected for this process. For each matched commit, we examined the changeset as well as up to four parent commits to identify any modifications to version-related files, such as the \codett{pom.xml} file for Maven artifacts. Our analysis revealed that 1,096 artifacts, or 1.5\% of the total sample, exhibited version mismatches in the files associated with the commit's source code. While this discrepancy rate is relatively low, offering some confidence in the accuracy of the matching process, it is important to note that manual verification is required to establish the ground truth. However, due to the scale of the analysis, such manual verification is not feasible.

\subsection{RQ2: Build Success on Reproducible Builds}

We evaluated the success of our tool in generating build specifications and rebuilding known reproducible packages. Specifically, we used the 100 packages from Reproducible Central, which were previously studied by AROMA~\cite{aroma}. These 100 packages, all in the Maven GAV format, served as the basis for our evaluation.

Out of the total packages, \tool successfully identified build commands and generated valid build specifications for 55 packages. For 19 packages, \tool was able to identify build commands from GitHub Actions Workflow files. However, none of these commands corresponded to either Maven or Gradle build tools, which are the primary tools supported by \tool. As a result, no build specifications could be generated for these packages. In the case of 26 packages, \tool was unable to find any build commands in the GitHub Actions Workflow files. To address this, we took a best-effort approach and used a default Maven build command~\footnote{mvn clean package -DskipTests=true -Dmaven.test.skip=true -Dmaven.site.skip=true -Drat.skip=true -Dmaven.javadoc.skip=true -Dgenerate-metadata=true} to generate the build specifications for these packages. These results indicate that \tool successfully identified build information for a significant portion of the Reproducible Central dataset, but there were some limitations, especially in cases involving unsupported build tools or missing build commands.

Additionally, we selected 19 Gradle-based packages from the Reproducible Central dataset to extend the evaluation beyond Maven-based projects supported by AROMA. These packages presented distinct challenges due to Gradle's different build mechanisms. For these 19 Gradle packages, we were able to successfully generate the build specifications and rebuild all of them. This demonstrates the broader applicability of \tool, as it extends beyond the Maven projects to handle Gradle projects, which have been overlooked in the previous study~\cite{aroma}.

\begin{table}[h!]
    \centering
    \caption{Rebuild status for 81 packages in the RC dataset, including additional 19 Gradle packages.}
    \label{tab:100_aroma_rc}
    \begin{tabular}{@{}cccccc@{}}
        \toprule
        RC & AROMA  & \tool  & Build Tool &  \# of Packages \\ \midrule
        FAILED               & FAILED & FAILED & Maven      & 3  \\
        PASSED               & PASSED & PASSED & Maven      & 73 \\
        FAILED               & PASSED & FAILED & Maven      & 5  \\
        PASSED               & \_     & PASSED & Gradle     & 19 \\ \midrule
                
    \end{tabular}
\end{table}

Table~\ref{tab:100_aroma_rc} presents the rebuild results for 81 packages from the Reproducible Central dataset, with 19 additional Gradle packages not covered by the AROMA dataset, and compares them against the outcomes produced by AROMA and the original Reproducible Central build specifications. Notably, \tool successfully rebuilt 73 out of 81 packages, demonstrating its robustness and effectiveness in handling a wide range of artifacts. Three packages could not be rebuilt using either the original Reproducible Central build specs or AROMA, with Java class loading errors. Additionally, for five packages where AROMA successfully rebuilt the artifacts, \tool encountered challenges. In all of these cases, \tool identified that the build had been executed using the Maven wrapper (\codett{mvnw}) as specified in the GitHub Actions configuration. However, for three of the packages, the Maven wrapper triggered an older version of Maven (v3.5.2), which led to compatibility issues with certain build plugins. The remaining two packages, both from the same project, faced compiler errors. In these instances, both the original Reproducible Central and AROMA, by default, used \codett{mvn} instead of the wrapper. Interestingly, attempting to reproduce the exact environment from GitHub Actions did not yield the expected results.

These observations highlight both the strengths and limitations of our tool in the context of known reproducible builds. While \tool performs well in many cases, specific challenges related to older build environments and project-specific build configurations still remain.

\subsection{RQ3: Build Success in the Wild}

In \cite{aroma}, the authors describe the selection of 100 packages with the \codett{project.build.outputTimestamp} property present in their POM files. Additionally, 100 packages were chosen that lacked this property, which was considered an indication of the project's non-compliance with reproducibility practices. These 200 packages were not part of the Reproducible Central dataset, ensuring that the selected artifacts were not part of previously examined reproducibility efforts.

In this section, first we evaluate \tool on these  200 packages. For 19 of these packages, \tool encountered challenges in generating valid Buildspec files due to specific characteristics of the projects. Of these packages, 13 were identified as webjars, which do not contain Java code, and thus could not be processed by \tool. One package used \codett{npx} to publish a JavaScript project as a JAR, which fell outside the scope of \tools build specification capabilities. Additionally, two packages had repository URLs that \tool could identify, but these were either nonexistent or incorrectly specified in the metadata, preventing the generation of build specifications. Despite these isolated cases, \tool successfully generated build specifications for the remaining 181 packages, demonstrating its robustness in handling a wide range of project configurations.

We proceeded to rebuild the 181 packages using RC for which valid build specifications had been generated. Although 30 packages were excluded from the rebuild due to taking more than one hour, 151 builds were successfully executed. Of these, 56 packages included the \codett{project.build.outputTimestamp} property in their POM files (Table~\ref{tab:56_aroma_non_rc_rebuild_with_output_timestamp}), while 70 packages did not ( Table~\ref{tab:70_aroma_non_rc_rebuild_no_output_timestamp}). These results showcase the high success rate of \tool in handling diverse package configurations.

\begin{table}[h!]
    \centering
    \caption{Rebuild status for 56 non-RC packages picked by AROMA that include the \codett{project.build.outputTimestamp} property in their POM file.}
    \label{tab:56_aroma_non_rc_rebuild_with_output_timestamp}
    \begin{tabular}{@{}cccc@{}}
    \toprule
    & AROMA                & \tool                &    \\ \midrule
    & FAILED               & PASSED               & 7  \\
    & PASSED               & PASSED               & 25 \\
    & PASSED               & FAILED               & 8 \\
    & FAILED               & FAILED               & 16 \\
    & \multicolumn{1}{l}{} & \multicolumn{1}{l}{} & 56 \\ \bottomrule
    \end{tabular}
\end{table}

\begin{table}[h!]
    \centering
    \caption{Rebuild status for 70 non-RC packages picked by AROMA that lack the \codett{project.build.outputTimestamp} property in their POM file.}
    \label{tab:70_aroma_non_rc_rebuild_no_output_timestamp}
    \begin{tabular}{@{}cccc@{}}
    \toprule
    & AROMA                & \tool                &    \\ \midrule
    & FAILED               & PASSED               & 2  \\
    & PASSED               & PASSED               & 43 \\
    & PASSED               & FAILED               & 3 \\
    & FAILED               & FAILED               & 22 \\
    & \multicolumn{1}{l}{} & \multicolumn{1}{l}{} & 70 \\ \bottomrule
    \end{tabular}
\end{table}

 Overall, \tool demonstrated a high success rate in rebuilding packages on both datasets. The comparison between \tool and AROMA demonstrates the robustness of \tool, which outperforms AROMA in several instances. We further analyzed the cases where \tool succeeded while AROMA failed. \tools build commands are more flexible and comprehensive than AROMA's, ensuring a higher likelihood of build success. \tool generates a more general build configuration, focusing on the entire project, which ensures that all sub-modules are considered during the build process. In contrast, AROMA uses the same build command for all projects and targets specific artifacts using the \codett{-pl} parameter, which limits the build scope to a particular artifact and its dependencies. This can result in build failures if the targeted artifact has unresolved dependencies or if the project configuration doesn't properly handle such targeted builds. Additionally, \tools broad build scope avoids the risk of build issues that can arise in large, complex multi-module projects. By building the entire project, \tool is less likely to encounter failures caused by missing or broken sub-modules, as it ensures that all necessary components are included.

In the 11 cases where \tool failed to rebuild the package but AROMA succeeded, the root causes can be attributed to specific issues. In one instance, the failure was due to an untagged build plugin in the \codett{pom.xml} file. The newer version of this build plugin was incompatible with the project, necessitating a patch to make the build process functional. For the remaining 10 cases, the failure was due to the unavailability or inability to resolve one or more dependencies, which were either no longer available on Maven Central or had been removed. This highlights a broader issue in the software supply chain: while a package may be successfully built at one point in time, it becomes impossible to rebuild if any of its dependencies become unavailable. Such issues are external to \tools functionality and reflect a common challenge faced by software ecosystems when dependencies are no longer maintained or accessible.~\footnote{https://github.com/jvm-repo-rebuild/reproducible-central/issues/662}\\

\begin{tcolorbox}[colframe=gray!75, colback=gray!10, coltitle=black, title=Key Insight]
\tool is able to automatically generate build specifications for the majority of artifacts in AROMA dataset. In cases where \tool faced difficulties while AROMA had succeeded previously, the root cause lies in a broader issue in the software supply chain: dependencies that were once available can suddenly become inaccessible, making it impossible to rebuild a package. This insight underscores a critical challenge in the software ecosystem: the fleeting nature of dependencies and their impact on long-term reproducibility.
\end{tcolorbox}

In the next experiemnt, we randomly selected 200 unique GAVs from the latest versions of the 1200 most popular packages in Maven Central (Same dataset used in Section~\ref{sec:motivation}). To ensure the robustness of our evaluation, we verified that none of these 200 packages were included in the rebuild datasets of either AROMA or Reproducible Central.

\begin{table}[ht]
\centering
\resizebox{\columnwidth}{!}{
\begin{tabular}{|l|c|}
\hline
\multicolumn{1}{|c|}{\textbf{Category / Finding}} & \textbf{Count} \\
\hline
\multicolumn{2}{|c|}{\textbf{Buildspec Generation Outcomes (200 Packages)}} \\
\hline
Buildspecs successfully generated & 134 \\
Buildspec generation pending improvement & 66 \\
\hline
\multicolumn{2}{|c|}{\textbf{Details of 66 Buildspec Generation Challenges}} \\
\hline
Non-Maven/Gradle builds (e.g., SBT, Ant, NPM/Yarn) & 23 \\
Lookup command parsing and patching issues & 7 \\
Repository information unavailable or inaccessible & 36 \\
\hline
\multicolumn{2}{|c|}{\textbf{Breakdown of Repository Discovery Challenges (36 Total)}} \\
\hline
Repository accessible, but tag could not be matched & 6 \\
Malformed SCM URL in \texttt{pom.xml} & 2 \\
SCM URL is a website rather than a VCS & 2 \\
Repository deleted or made private & 12 \\
No SCM metadata in \texttt{pom.xml} & 7 \\
Unsupported or unreachable git hosting services & 6 \\
False negative in repository detection logic & 1 \\
\hline
\multicolumn{2}{|c|}{\textbf{Rebuild Outcomes for Successfully Generated Buildspecs (134 Total)}} \\
\hline
Successful Gradle rebuilds & 10 \\
Successful Maven rebuilds & 30 \\
Gradle rebuilds with errors & 34 \\
Maven rebuilds with errors & 35 \\
Builds with system-level issues (e.g., environment or setup) & 25 \\
\hline
\multicolumn{2}{|c|}{\textbf{System-Level Issues (25 Total)}} \\
\hline
Critical JVM-related errors & 3 \\
Gradle execution issues (RC environment-specific) & 13 \\
Maven execution issues (RC environment-specific) & 2 \\
Unsupported JDK version (e.g., JDK 22-ea) & 1 \\
Build interrupted or incomplete & 2 \\
Spec mistakenly used an unsupported tool (e.g., not Maven/Gradle) & 4 \\
\hline
\textbf{Total Packages Evaluated} & \textbf{200} \\
\hline
\end{tabular}
}
\caption{Insights from Evaluating 200 Maven Packages not covered by AROMA and Reproducible Central}
\label{tab:buildgen_summary}
\end{table}

Table \ref{tab:buildgen_summary} provides an overview of the build specification generation and rebuild outcomes for 200 randomly selected Maven packages not previously evaluated by AROMA or Reproducible Central. \tool demonstrated strong performance, successfully generating build specifications for 134 packages, with challenges mainly stemming from non-Maven/Gradle builds and issues with repository information. Notably, \tools comprehensive approach allows for detailed insights into build failures and system-level issues, such as Gradle and Maven execution problems, highlighting areas where further improvements can be made.

In cases where the build had failed, we sought to determine whether the failures were attributed to the unique setup of Reproducible Central, which installs specific build tools and enforces predefined configurations. The results, presented in Table~\ref{tab:build-outside-rc}, were rather surprising. A considerable number of the artifacts were successfully built outside of Reproducible Central, on a virtual machine equipped with all major Java versions, using the same build commands identified by \tool. Of the 25 system-level issues encountered in Table~\ref{tab:buildgen_summary}, 13 builds were successfully completed, representing a 52\% improvement. Additionally, among the 69 failed Maven and Gradle builds, we were able to rebuild 20 packages (17 Gradle and 3 Maven), yielding a 28.9\% improvement. Further investigation revealed that the Java version was incorrectly specified in some builds. By correcting the Java version based on error logs, we managed to rebuild 25 more packages (19 Gradle and 6 Maven), yielding a 36.2\% improvement.

These findings underscore the importance of log analysis and establishing a feedback loop to further improve \tools performance. The results also reveal that the rebuild infrastructure plays a pivotal role in the success of the build process, highlighting the need for improvements in this area to fully leverage \tools potential. A promising avenue for future development is enhancing the reliability of the underlying build infrastructure. In this context, exploring projects like OSS Rebuild~\footnote{https://github.com/google/oss-rebuild}, which could provide an alternative infrastructure for running \tools automatically generated \spec, could offer valuable insights and further boost \tools effectiveness.\\

\begin{tcolorbox}[colframe=gray!75, colback=gray!10, coltitle=black, title=Key Insight]
Our findings emphasize the critical role of log analysis and a feedback loop in refining \tools performance. Furthermore, the rebuild infrastructure significantly impacts build success, highlighting the need for improvements in RC to maximize \tools potential.

\end{tcolorbox}

\begin{table*}[ht]
\centering
\caption{Running \tools \spec without RC Infrastructure, with Improvements in Success Count}
\label{tab:build-outside-rc}
\begin{tabular}{|l|c|c|c|}
\hline
\textbf{Category} & \textbf{Failure Count with RC} & \textbf{Fail Count without RC} & \textbf{Improvement (\%)} \\ \hline
\textbf{System-Level Issues} & 25 & 12 & 52\% \\ \hline
\textbf{Other Failed Maven/Gradle Builds} & 69 & 49 & 28.9\% \\ \hline
\end{tabular}
\end{table*}

\section{Related Work}

This section provides related research on the challenges of rebuilding artifacts and the security risks posed by artifacts that cannot be built from source.

\paragraph{Security and Risk Management via Rebuilding Artifacts}
The ability to rebuild open-source artifacts is not only a software engineering challenge but also a significant security concern. Research has demonstrated that unreproducible builds can increase the risk of malicious attacks. For instance, \cite{Backstabber-20dimva, hercule} investigates real-world incidents involving malicious packages injected into open-source repositories like npm, PyPI, and RubyGems. These attacks exploit the non-deterministic nature of builds, allowing malware to be injected during package installation. These studies emphasize the importance of isolating and hardening the build process to mitigate such security threats.

Building on this, SRZ et al.~\cite{SRZ-23bigdata} propose a comprehensive risk model that integrates build environment details, build steps, and dependencies. This framework allows organizations to assess and manage risks within their software supply chains more effectively. In parallel, the significance of Software Bill of Materials (SBOMs) in identifying third-party components and their dependencies is highlighted in the works of \cite{LandscapeSBOM-2024} and \cite{SWC-24icse}. SBOMs play a crucial role not only in vulnerability management but also in ensuring reproducibility across different infrastructures. By resolving dependencies deterministically, SBOMs help ensure that the rebuild process can be reliably executed, regardless of the environment.

These efforts underscore the importance of deterministic builds in securing the software supply chain, aligning closely with our goal of improving rebuild processes through automated specification generation.

\paragraph{Challenges and Limitations of Reproducible Builds}
Reproducible builds and their associated environments have garnered significant attention in both open and closed-source software development. A key study in this area~\cite{business_adoption} explores the challenges and benefits businesses encounter when adopting reproducible build practices, offering valuable insights into industry perspectives. Similarly, research on mitigating the risks associated with unreproducible builds underscores the security implications of non-reproducibility in the software supply chain~\cite{reducing_risks}. Drawing on examples from high-profile attacks like SolarWinds and Log4j, these findings highlight the critical importance of reproducible builds to prevent malicious exploits.

Further research has delved into the role of packaging and distribution in open-source ecosystems. An empirical study on reproducible packaging~\cite{reproducible_packaging} investigates how packaging systems can be optimized to support reliable and reproducible builds. Concurrently, work on the preservation of build environments~\cite{MZZ-24icse} emphasizes that maintaining both the source code and the environment is essential for achieving deterministic builds. In the \textit{npm} ecosystem, Goswami et al.~\cite{goswami2020investigating} identify challenges related to tools like \textit{UglifyJS} and \textit{Babel}, which introduce variability in build outcomes, further complicating efforts to ensure reproducibility.

Despite these advancements, several challenges persist. \cite{FWE-23SandP} discusses the inherent limitations of achieving perfect reproducibility, noting that factors such as timestamps, file paths, and concurrency introduce non-determinism, which complicates the goal of full reproducibility. The study also highlights that many users are satisfied with ``reproducible-ish'' builds, especially when they do not perceive any immediate security threats from unreproducibility. Similarly, \cite{ZGF-23actionsremaker} introduces \textit{ActionsRemaker}, a system designed to reproduce GitHub Action workflows using Docker containers. However, the study acknowledges the challenge of ensuring that all required dependencies are correctly installed, which can result in inconsistencies in the build process. Furthermore, research by Bajaj et al.~\cite{bajaj2024unreproducible} identifies system architecture differences and locale configurations as significant contributors to build unreproducibility, highlighting the complexity of achieving fully deterministic builds across diverse environments. 

These challenges point to the need for addressing challenges to improve rebuilding artifacts. Motivated by these issues, our work seeks to improve the process of rebuilding artifacts by reducing the manual effort involved in collecting build information. We leverage static analysis techniques to automate the extraction of critical build information, specifically within the Maven ecosystem supporting popular tools, such as Maven and Gradle. By conducting large-scale analysis, we automate the identification of source code and commit versions, ultimately generating precise build specifications. This approach not only addresses the technical complexities and environmental variations highlighted in prior research but also reduces uncertainty and minimizes manual intervention, making the process of rebuilding artifacts more automatic.

\subsection{Rebuilding Maven Artifacts for Testing}

Several projects have explored automated build systems that execute test suites, such as BugSwarm~\cite{bugswarm}, GitBug-Java~\cite{gitbug}, and ExecutionAgent~\cite{llmagent}. However, these systems generally do not support source code identification, certain CI configurations (e.g., GitHub Actions), or occasionaly require paid LLMs to function effectively.

Building on the Macaron~\cite{macaron} framework, a robust industry-grade solution for supply chain security, our work enhances its functionality to automate the rebuilding of open-source projects. Unlike existing systems, our approach simplifies the process by requiring only the PackageURL identifier of an artifact. The subsequent tasks, including source code detection and \spec generation, are fully automated, with no additional cost, making the entire process both efficient and accessible.

\paragraph{Analysis of unreproducible builds}
The challenge of unreproducible builds has motivated the development of various approaches for identifying the causes of non-reproducible behavior. One early approach focuses on localizing discrepancies in the build environment that contribute to unreproducibility~\cite{automated_localization}. More advanced techniques, such as causality analysis over system call traces~\cite{root_cause_localization}, analyze interactions between system calls to pinpoint the root causes of non-reproducibility. In addition to detection, automated patching methods~\cite{automated_patching} have been proposed to automatically apply fixes that mitigate sources of unreproducibility, improving the overall build environment. Another debugging technique leverages eBPF (extended Berkeley Packet Filter)~\cite{debugging_ebpf}, which offers detailed insights into the execution flow of build processes, enabling the identification of specific steps where reproducibility fails. These methods are powerful tools in diagnosing and resolving unreproducibility, but they often assume that the build process has already been successfully captured or replicated.

In contrast to these approaches, we propose an automated system that generates build specifications for rebuilding Maven artifacts directly from source. Our method provides a proactive solution: rather than merely diagnosing issues post-failure, \tool enables users to reproduce builds from the ground up, ensuring greater initial reproducibility. Furthermore, when builds do fail, our approach can complement existing localization techniques by providing a consistent starting point, thus facilitating the application of tools like system call tracing or eBPF for deeper diagnostics. In this way, \tool not only addresses the problem of reproducibility but also integrates with other debugging strategies to enhance overall effectiveness in tackling unreproducible builds.

\section{Conclusion}
\label{conclusion}

Maven Central plays a critical role in modern software supply chains, but the separation between binaries and source code creates significant transparency and security challenges. With a large number of top Maven artifacts lacking transparent CI/CD pipelines, users must place considerable trust in both source code and build environments.

Rebuilding artifacts from source improves supply chain security by enabling code review, binary-source verification, and control over dependencies. However, challenges arise due to variations in build environments. In this paper, we present an extension to Macaron, an open-source framework, that automates the rebuilding of Maven artifacts. Our approach enhances performance in source code detection, automates the extraction of build specifications from GitHub Actions, and provides a scalable solution to rebuild artifacts, thus improving security and transparency in open-source supply chains.  Furthermore, we provide a thorough root cause analysis of build failures in Java projects, laying the foundation for future research to improve the success rate of rebuilding artifacts.

\bibliographystyle{plain}
\bibliography{refs}

\end{document}